\documentclass[12pt]{article}

\usepackage{eqnarray,amsmath}
\usepackage{amssymb}
\usepackage{cancel}
\usepackage{xfrac}

\usepackage{fourier}
\usepackage{ulem}
\usepackage[T1]{fontenc}

\usepackage{accents}
\usepackage{sectsty}
\sectionfont{\fontsize{12}{15}\selectfont}
\subsectionfont{\normalfont\fontsize{12}{15}\selectfont}
\subsubsectionfont{\itshape\normalfont\fontsize{12}{15}\selectfont}
\usepackage{bm}
\usepackage{lineno}

\usepackage{relsize}

\DeclareMathOperator{\expit}{expit}

\usepackage{accents}
\newlength{\dhatheight}

\newcommand*\patchAmsMathEnvironmentForLineno[1]{%
	\expandafter\let\csname old#1\expandafter\endcsname\csname #1\endcsname
	\expandafter\let\csname oldend#1\expandafter\endcsname\csname end#1\endcsname
	\renewenvironment{#1}%
	{\linenomath\csname old#1\endcsname}%
	{\csname oldend#1\endcsname\endlinenomath}%
}
\newcommand*\patchBothAmsMathEnvironmentsForLineno[1]{%
	\patchAmsMathEnvironmentForLineno{#1}%
	\patchAmsMathEnvironmentForLineno{#1*}%
}
\AtBeginDocument{%
	\patchBothAmsMathEnvironmentsForLineno{equation}%
	\patchBothAmsMathEnvironmentsForLineno{align}%
	\patchBothAmsMathEnvironmentsForLineno{flalign}%
	\patchBothAmsMathEnvironmentsForLineno{alignat}%
	\patchBothAmsMathEnvironmentsForLineno{gather}%
	\patchBothAmsMathEnvironmentsForLineno{multline}%
}

\usepackage{color}
\usepackage{wrapfig}
\usepackage{indentfirst} 
\usepackage{setspace}
\usepackage[numbers,super,sort&compress]{natbib} 
\usepackage{booktabs} 
\usepackage{rotating} 
\usepackage[margin = .85in]{geometry} 
\usepackage{fancyhdr} 
\usepackage{tabularx}
\usepackage{pdfpages}
\usepackage{longtable}

\usepackage{subfig}
\usepackage{mdwlist}
\usepackage{url}
\usepackage{verbatim}
\usepackage{multirow}
\usepackage{multicol}
\usepackage[compact]{titlesec}
\usepackage[colorlinks = TRUE, urlcolor = blue, linkcolor = black, citecolor = black]{hyperref}
\usepackage{float}
\usepackage{lscape}
\usepackage{graphicx}
\makeatletter
\renewcommand\@biblabel[1]{#1.}
\makeatother

\begin{document}
\thispagestyle{empty}
\baselineskip=28pt

\vskip .2cm
\begin{center}
{{\noindent \Large Challenges in Obtaining Valid Causal Effect Estimates with Machine Learning Algorithms}}\\
Ashley I. Naimi, Alan E. Mishler, Edward H. Kennedy\footnote{Ashley Naimi is Assistant Professor in the Department of Epidemiology, University of Pittsburgh. Alan Mishler is a Doctoral Student in the Department of Statistics \& Data Science, Carnegie Mellon University. Edward Kennedy is Assistant Professor in the Department of Statistics \& Data Science, Carnegie Mellon University. The authors are indebted to Drs. Stephen R. Cole, Michael Hudgens, Alex Breskin, and Paul Zivich (UNC) for helpful comments and expert advice. This research was supported in part by the University of Pittsburgh Center for Research Computing through the computing resources provided, and the assistance of Dr. Kim Wong. This work was funded by the NIH grant number R01HD093602.}
\end{center}

\baselineskip=12pt

\begin{center}
{{\bf Abstract}}
\end{center}
\baselineskip=12pt
Unlike parametric regression, machine learning (ML) methods do not generally require precise knowledge of the true data generating mechanisms. As such, numerous authors have advocated for ML methods to estimate causal effects. Unfortunately, ML algorithms can perform worse than parametric regression. We demonstrate the performance of ML-based single- and double-robust estimators. We use 100 Monte Carlo samples with sample sizes of 200, 1200, and 5000 to investigate bias and confidence interval coverage under several scenarios. In a simple confounding scenario, confounders were related to the treatment and the outcome via parametric models. In a complex confounding scenario, the simple confounders were transformed to induce complicated nonlinear relationships. In the simple scenario, when ML algorithms were used, double-robust estimators were superior to single-robust estimators. In the complex scenario, single-robust estimators with ML algorithms were at least as biased as estimators using misspecified parametric models. Double-robust estimators were less biased, but coverage was well below nominal. The use of sample splitting, inclusion of confounder interactions, reliance on a richly specified ML algorithm, and use of doubly robust estimators was the only explored approach that yielded negligible bias and nominal coverage. Our results suggest that ML based singly robust methods should be avoided.
%
%
%
\baselineskip=12pt
\vskip 1cm \par\noindent
\par\medskip\noindent

%
%
\newpage
\thispagestyle{empty}

\newpage
\doublespacing
\setcounter{page}{1}

Both machine learning methods and doubly robust estimators are becoming increasingly popular, yet the critical relation between them remains poorly understood. Machine learning methods consist of a wide range of analytic techniques that do not require hard to verify modeling assumptions. Because of this, they are often assumed to be less biased than their standard parametric counterparts. This perceived property has motivated many to either recommended or use machine learning methods to quantify exposure effects.\cite{Lee2010,Westreich2010c,Snowden2011,Oulhote2019} However, it is generally not recognized that machine learning methods can yield effect estimates that are more biased, with poorer confidence interval coverage, than their parametric counterparts. These problems arise due to the curse of dimensionality.\cite{Chernozhukov2018,Hastie2009,Naimi2020}

Doubly robust estimators are so named because these methods allow two chances for adjustment.\cite{Robins1995b,Robins2001a,Bang2005} In the case of confounding adjustment, these chances arise because the analyst must fit two models: a model for the outcome regressed against the exposure and all confounders (outcome model); and a model for the exposure regressed against all confounders (the propensity score model). These are then combined to estimate the effect of interest.\cite{Rotnitzky2014} 

The benefits of doubly robust methods have been explained by pointing out that if a confounding variable is left out of either the exposure or the outcome model (but not both), unbiased estimates can still be obtained.\cite{Jonsson-Funk2011} While true, analysts would not typically leave confounding variables out of either the exposure or outcome model. Such justifications ignore a critically important benefit conferred by doubly robust estimators: under relatively mild conditions, they remain unbiased, with asymptotically nominal confidence interval coverage, even when machine learning methods are used to fit the exposure and outcome models.\cite{vanderLaan2006,Kennedy2017} In effect, doubly robust methods can mitigate or resolve problems caused by the curse of dimensionality.

This little recognized relation between machine learning and doubly robust estimators has important implications for applied researchers, particularly those interested in using machine learning methods to estimate causal effects. Here, we examine these implications using Monte Carlo simulations.\cite{Metropolis1949} Our intent is to clarify that machine learning methods should be used with doubly robust methods; they should not generally be used to estimate causal effects with singly robust techniques, such as model-based standardization, or inverse probability weighting.

\section*{Observed Data \& Target Parameter}

We consider a simple setting with a single binary exposure ($X$), a set of continuous confounders ($\mathbf{C} = \{C_1,C_2, C_3, C_4\}$) measured at baseline, and a single continuous outcome ($Y$) measured at the end of follow-up. In an observational cohort study to estimate the effect of $X$ on $Y$, $\mathbf{C}$ might be assumed a minimally sufficient adjustment set,\cite{Greenland1999a} and the exposure and outcome would be assumed generated according to some unknown models, for example:
\begin{align}
		& P(X = 1 \mid \mathbf{C}) = f(\mathbf{C}) \label{propensity} \\
		& E(Y \mid X, \mathbf{C}) = g(X,\mathbf{C}), \label{outcome}
\end{align}
where $f(\bullet)$ and $g(\bullet)$ represent functions of $C$, and $X$ and $C$, respectively. In an observational cohort study assuming a correct confounder adjustment set, this is the extent of what is known about the exposure and outcome models.\cite{Robins2001} 

We focus here on the average treatment effect:
\begin{equation*}
	\psi = E(Y^{x=1} - Y^{x=0}) 
\end{equation*}
where $Y^x$ is the outcome that would be observed if $X$ were set to $x$. This estimand is (point) identified by
$$ \psi = E\{ g(X=1,\mathbf{C})-g(X=0,\mathbf{C})\} = E\left \{ \left [ \frac{XY}{f(\mathbf{C})} \right ] -  \left [ \frac{(1-X)Y}{1-f(\mathbf{C})} \right ]\right \} $$ 
under positivity, consistency, and exchangeability.\cite{Robins2009,Naimi2016b} If these assumptions hold, $\psi$ can be estimated using a number of approaches. In the equations that follow, we let $i$ index sample observations, and $\hat{f}_i(\mathbf{C})$ and $\hat{g}_i(X=x,\mathbf{C})$ are individual sample predictions for $P(X = 1 \mid \mathbf{C})$ and $E(Y \mid X=x,\mathbf{C})$, respectively.

With predictions from Model \ref{propensity}, $\psi$ can be estimated via inverse probability weighting\cite{Hernan2006} as:
\begin{equation}
	\hat{\psi}_{ipw} = \frac{1}{N} \sum_{i = 1}^N \left \{ \left [ \frac{X_iY_i}{\hat{f}_i(\mathbf{C})} \right ] -  \left [ \frac{(1-X_i)Y_i}{1-\hat{f}_i(\mathbf{C})} \right ]\right \}. \label{ipw}
\end{equation} 
With predictions from Model \ref{outcome}, $\psi$ can be estimated via model-based standardization (henceforth g computation)\cite{Naimi2016b}:
\begin{equation}
	\hat{\psi}_{gComp} = \frac{1}{N}\sum_{i=1}^N \big \{ \hat{g}_i(X=1,\mathbf{C}) - \hat{g}_i(X=0,\mathbf{C}) \big \}. \label{gComp}
\end{equation}
Both approaches \ref{ipw} and \ref{gComp} are ``singly robust'' in that they typically rely entirely on the correct specification of the appropriate single regression model. If these models are misspecified, the estimators will not generally converge to the true value. 

Alternatively, one may employ a ``doubly robust'' technique where predictions from both the exposure and outcome models are combined into a single estimator to quantify the effect of interest. For example, using predictions from both Models \ref{propensity} and \ref{outcome}, $\psi$ can be estimated as:
\begin{equation}
\hat{\psi}_{aipw} = \frac{1}{N}\sum_{i=1}^N \left \{ \frac{(2X_i-1)[Y_i - \hat{g}_i(X,\mathbf{C})]}{(2X_i-1)\hat{f}_i(\mathbf{C}) + (1-X_i)} + \hat{g}_i(X=1,\mathbf{C}) - \hat{g}_i(X=0,\mathbf{C}) \right \}. \label{aipw}
\end{equation}
Equation \ref{aipw} is an augmented inverse probability weighted estimator, and will converge to the true value as the sample size grows if either $f(\mathbf{C})$ or $g(X,\mathbf{C})$, but not necessarily both, are consistently estimated. The estimator \ref{aipw} can be viewed as either a bias-corrected version of the g computation estimator (where the correction is the term incorporating the propensity score defined in \ref{propensity}), or an efficiency enhanced version of the IPW estimator (where the enhancement is the term incorporating the outcome model defined in \ref{outcome}).\cite{Daniel2018}

Alternatively, model \ref{propensity} can be used to ``update'' model \ref{outcome} via targeted minimum loss-based estimation:\cite{Rose2011}$^{(p72-3)}$
\begin{equation}
	\hat{\psi}_{tmle} = \frac{1}{N}\sum_{i=1}^N \big \{ \hat{g}^{u}_i(X=1,\mathbf{C}) - \hat{g}^{u}_i(X=0,\mathbf{C}) \big \}, \label{tmle}
\end{equation}
where $\hat{g}^{u}_i(X=1,\mathbf{C})$ are predictions from an ``updated'' outcome model. For the average treatment effect, this outcome model is updated by first generating an inverse probability weight, defined as:
 \[
    H(X,\mathbf{C})=\left\{
                \begin{array}{ll}
                  \frac{1}{\hat{f}_i(\mathbf{C})} & \text{if }X=1 \\
                  - \frac{1}{1-\hat{f}_i(\mathbf{C})} & \text{otherwise}
                \end{array}
              \right.
  \]
and then including this inverse probability weight in a no-intercept logistic regression model for the outcome that includes the previous outcome predictions $\hat{g}_i(X,\mathbf{C})$ as an offset. The $\hat{g}^{u}_i(X=1,\mathbf{C})$ predictions are then generated from this model by setting $X$ to 1 and then to 0 for all individuals in the sample. TMLE is asymptotically equivalent to equation \ref{aipw} but can have better finite-sample performance since the resulting estimate will be appropriately bounded by, e.g., the minimum and maximum empirical values of $Y$.\cite{Gruber2012}

\section*{Parametric Estimation}

For binary $X$ and continuous $Y$, it is customary to specify models \ref{propensity} and \ref{outcome} parametrically using logistic and linear regression, respectively:
\begin{align}
		& P(X = 1 \mid \mathbf{C}) = \expit(\alpha_0 + \alpha_1C_1 + \alpha_2C_2 + \alpha_3C_3 + \alpha_4C_4), \label{parm_propensity}\\ & \hskip 3cm  \expit(\bullet) = 1/(1+\exp[-\bullet])  \notag \\
		& E(Y \mid X, \mathbf{C}) = \beta_0 + \beta_1 X + \beta_2 C_1 + \beta_3 C_2 + \beta_4 C_3 + \beta_5 C_4, \label{parm_outcome}\\& \hskip 3cm Y \mid X, \mathbf{C} \sim \mathcal{N}\Big(E(Y \mid X, \mathbf{C}),\sigma^2 \Big) \notag 
\end{align}
where we let $\beta_0 + \beta_1 X + \beta_2 C_1 + \beta_3 C_2 + \beta_4 C_3 + \beta_5 C_4 = \mu$, and we collectively refer to all the $\beta$'s in model \ref{parm_outcome} as $\boldsymbol{\beta}$. Imposing these forms on $f(\mathbf{C})$ and $g(X,\mathbf{C})$ permits use of standard maximum likelihood for estimation and inference.\cite{Cole2013a}

\subsection*{\textit{Estimation via Parametric Outcome Model}}

Model \ref{parm_outcome} imposes several parametric constraints on the form of $g(X, \mathbf{C})$: (i) $Y$ follows a conditional normal distribution with constant variance not depending on $X$ or $\mathbf{C}$; and (ii) the mean $\mu$ is related to the covariates $X$ and $\mathbf{C}$ additively, as detailed in model \ref{parm_outcome}. If these constraints on $g(X,\mathbf{C})$ are true, and other identification and regularity conditions hold,\cite{Longford2008}$^{(ch2)}$ the maximum likelihood estimates of $\boldsymbol{\beta}$ are asymptotically efficient.\cite{Rencher2000}$^{(p144)}$ Relatedly, under the model constraints and identification and regularity conditions, as the sample size increases, the estimates of $g(X,\mathbf{C})$ and/or $f(\mathbf{C})$ will converge to the true values at an optimal (i.e., $\sqrt{N}$) rate, and their distribution will be such that confidence intervals can be easily derived.

If constraint (i) is violated, the maximum likelihood estimator is no longer the most efficient, but can still be used to estimate $\psi$ consistently. If constraint (ii) is violated, then the maximum likelihood estimator is no longer consistent. Depending on the severity to which constraint (ii) is violated, the bias may be substantial. Unfortunately, in an observational study the true form of model \ref{parm_outcome} is almost never known. This means that such maximum likelihood estimates are almost always biased, with the degree of bias depending on the (unknown) extent to which the model is mis-specified.\cite{Box1976}

\subsection*{\textit{Estimation via Parametric Exposure Model}}

One way to avoid relying on correct outcome model specification is to use a parametric model for exposure model \ref{propensity}, and estimate $\psi$ via $\hat{\psi}_{ipw}$. Specifically, with IP-weighting, one need not model the interactions between the exposure and any covariates.\cite{Hernan2001} Such an estimator is not as efficient as $\hat{\psi}_{gComp}$, and can be subject to important finite-sample biases when weights are very large. But as the sample size increases, the inverse probability weighted estimator converges at the same standard $\sqrt{N}$ rate as the g computation estimator.\cite{Westreich2012a} Unfortunately, as with the outcome model, the true form of model \ref{propensity} will almost never be known in an observational study. Mis-specification of model \ref{parm_propensity} will also lead to biased estimation of $\psi$, again with the degree of bias depending on the unknown extent of model mis-specification.

\subsection*{\textit{Parametric Doubly Robust Estimation}}

To mitigate against mis-specification of the exposure or outcome models, numerous authors have advocated for the use of estimators such as equation \ref{aipw} or \ref{tmle}. These doubly robust estimators remain consistent even if either the exposure model or the outcome model is mis-specified, but not both. However, if it is unlikely that either model \ref{parm_propensity} or \ref{parm_outcome} is correct, then the doubly robust estimator will also likely be biased, and not much better than the singly robust estimators.\cite{Kang2007, Kennedy2017}

\section*{Nonparametric Singly Robust Estimation: The Curse of Dimensionality}

Nonparametric methods are an alternative to parametric models. For example, nonparametric maximum likelihood estimation (NPMLE) for models \ref{propensity} or \ref{outcome} would entail fitting models \ref{parm_propensity} and \ref{parm_outcome}, but with a parameter for each unique combination of values defined by the cross-classification of all covariates (i.e., saturating the model). However, the NPMLE will be undefined in any finite sample with a continuous confounder, since there will be no covariate patterns containing both treated and untreated subjects. 

Alternatively, one can use nonparametric ``machine learning'' methods like kernel regression, splines, random forests, boosting, etc., which exploit smoothness across covariate patterns to estimate the regression function. However, for any nonparametric approach there is an explicit bias-variance trade-off that arises in the choice of tuning parameters; less smoothing yields smaller bias but larger variance, while more smoothing yields smaller variance but larger bias (parametric models can be viewed as an extreme form of smoothing). This tradeoff has important consequences. In particular, it is generally impossible to estimate regression functions at the standard $\sqrt{N}$ rates attained by correctly specified parametric estimators.\cite{vanderVaart2000} The consequence of these slower than optimal convergence rates is increased bias, and confidence interval estimators with less than nominal coverage.
 
Convergence rates for nonparametric estimators become slower with more flexibility and more covariates. For example, a standard rate for estimating smooth regression functions is $N^{-\beta/(2\beta+d)}$, where $\beta$ represents the number of derivatives of the true regression function, and $d$ represents the dimension of, or number of covariates in, the true regression function. This issue is known as the curse of dimensionality.\cite{Gyorfi2002,Robins1997c,Wasserman2006} Sometimes this is viewed as a disadvantage of nonparametric methods; however, it is just the cost of making weaker assumptions: if a parametric model is misspecified, it will converge very quickly to the wrong answer. 

In addition to slower convergence rates, confidence intervals are harder to obtain. Specifically, even in the rare case where one can derive asymptotic distributions for nonparametric estimators, it is typically not possible to construct confidence intervals (even via the bootstrap) without impractically undersmoothing the regression function (i.e., overfitting the data).\cite{Wasserman2006} 

These complications (slow rates and lack of valid confidence intervals) are generally inherited by the singly robust estimators \ref{ipw} and \ref{gComp} (apart from a few special cases which require simple estimators, such as kernel methods with strong smoothness assumptions and careful tuning parameter choices that are suboptimal for estimating $f$ or $g$).  For general nonparametric estimators $\hat{f}$ and $\hat{g}$, the estimators \ref{ipw} and \ref{gComp} will converge at slow rates, and honest confidence intervals will not be computable.

\section*{Nonparametric Doubly Robust Estimation}

Fortunately, doubly robust estimators that rely on nonparametric estimates of $f$ and $g$ do not suffer from the same limitations as the nonparametric versions of the singly robust estimators. In particular the doubly robust estimators \ref{aipw} and \ref{tmle} can be $\sqrt{N}$-consistent, asymptotically normal, and optimally efficient even if the estimators $\hat{f}$ and $\hat{g}$ are converging at slower nonparametric rates. In other words, the doubly robust estimator is less susceptible to the curse of dimensionality. This is a result of the fact that the error of the doubly robust estimator depends on the \textit{product} of the errors of $\hat{f}$ and $\hat{g}$, which goes to zero as fast or faster than either error alone. In particular, if $\hat{f}$ and $\hat{g}$ are converging to their targets at least faster than $n^{-1/4}$ rates (in $L_2$ norm), the doubly robust estimator will behave asymptotically just as if both $f$ and $g$ were estimated with correct parametric models. Importantly, $n^{-1/4}$ rates can be attained nonparametrically under relatively weak smoothness, sparsity, or other structural assumptions.\cite{Gyorfi2002, Wasserman2006} This improved performance of nonparametric methods when used with doubly robust techniques has important implications for applied researchers.

\section*{Simulation Study}

\subsection*{\textit{Data Generating Mechanism: Correct Specification}}

To explore these implications, we carried out a simulation study of singly and doubly robust estimators with parametric and nonparametric methods. We simulated 100 Monte Carlo samples, with sample sizes of \{200, 1200, 5000\} using data generating mechanisms that would lead to both simple and challenging conditions for estimation and inference. Specifically, we generated four independent standard normal confounders, denoted $C$. Both the exposure and outcome models included each of these confounders. The exposure was generated from a logistic model with: 
\begin{equation}
	P ( X = 1 \mid C) = \expit \left \{-1+\log(1.75)C_1+\log(1.75)C_2+\log(1.75)C_3+\log(1.75)C_4\right \},
\end{equation}
A continuous outcome was generated as:
\begin{equation}
\begin{split}
	Y = 120 + 6 X + 3 C_1 + 3 C_2 + 3 C_3 + 3 C_4 + \epsilon,
	\label{outcome_sim}
\end{split}	
\end{equation}
where the \underline{true average treatment effect $\psi = 6$}, with $\epsilon$ drawn from a normal distribution with mean $\mu=0$ and standard deviation $\sigma=6$. 

\subsection*{\textit{Data Generating Mechanism: Model Misspecification}}

To induce model misspecification, we followed previous research\cite{Kang2007} and transformed each of the continuous confounders as follows:
\begin{align*}
  Z_1 = \exp(C_1/2)\\
  Z_2 = C_2/(1+\exp(C_1))+10\\
  Z_3 = (C_1C_3/25+0.6)^3\\
  Z_4 = (C_2+C_4+20)^2
\end{align*}
Thus, while the true models generating the exposure and outcome variables included only the untransformed variables $C$, analyses conducted under parametric model misspecification included only the transformed variables $Z$.

\subsection*{\textit{Simulation Analysis}}

In each Monte Carlo sample, we estimated the average treatment effect $\psi = E(Y^1 - Y^0) = 6$ using g computation, inverse probability weighting, augmented inverse probability weighting, and targeted minimum loss-based estimation under two settings: ($i$) only the simple confounder data $C$ were available and used to specify all models (parametric and nonparametric), and ($ii$) only the transformed confounder data $Z$ were available and used to specify all models (parametric and nonparametric).

Parametric models were implemented as generalized linear models, with a binomial distribution and logistic link for the exposure, and a Gaussian distribution and identity link for the outcome model. As described above, these parametric models are correctly specified when the simple confounders are used, but highly misspecified when the transformed confounders are used.

Nonparametric estimation was accomplished via a stacking algorithm (Super Learner).\cite{vanderLaan2007} To explore the importance of the selected algorithm, we implemented a wide variety of different stacking algorithms that included different sets of base algorithms. Full details on all variations of the stacking algorithms explored are available in the Online Web Supplement. Here, we focus on stacked generalizations that included:
\begin{itemize}
	\item[version 1)] ($i$) random forests with 500 trees, random subspace selection value of two, and a minimum node size of 30 and 60; ($ii$) the extreme gradient boosting algorithm with 500 trees, a maximum tree depth of 4, shrinkage parameter of 0.1, and minimum node size of 30 and 60.
	\item[version 2)] Both random forests and extreme gradient boosting included in version 1, as well as ($iii$) generalized additive models with univariate smoothing splines with effective degrees of freedom between 3 and 8.
\end{itemize}

We also explored estimating the average treatment effects of interest with the stacking algorithms in version 2 that included 2-way interactions between all four confounders in the adjustment set. For all stacking algorithms, cross validation was used to compute the learner weights with fold sizes of $K = 10, 5,$ and $5$ for the sample sizes 200, 1200, and 5000, respectively.\cite{Naimi2018} For each machine learning based doubly robust estimator, we also explored the impact of sample splitting.\cite{Rinaldo2018,Zivich2020} This procedure involves splitting the sample into $K$ equal size folds, fitting models for $f(\mathbf{C})$ and $g(X,\mathbf{C})$ in one fold, using these models to predict exposure and outcome values in all remaining folds, and then repeating the process with the folds switched. The final effect estimate is computed over the entire sample as usual.

Standard errors for g computation were obtained from the standard deviation of 100 bootstrap resamples. However, for computational reasons, we were only able to apply the bootstrap to the nonparametric g computation estimator in select scenarios (see Online Web Supplement). Standard errors for the inverse probability weighted approach were obtained using the robust variance estimator. Standard errors for both doubly robust approaches were obtained using the variance of the efficient influence function. All confidence intervals were computed via the normal interval (i.e., Wald) equation. For each estimator in each scenario, we computed the bias: $B(\hat{\psi}) = E(\hat{\psi}) - \psi $, and 95\% confidence interval coverage, defined as the proportion of 95\% confidence intervals that included the true value over all 200 Monte Carlo runs. Simulations were done in \texttt{R} version 3.6.1. Code to reproduce our results is available on \href{https://github.com/ainaimi/NPDR}{GitHub}.

\section*{Simulation Results}

\begin{figure}[H]\rule{\textwidth}{1pt}
\begin{center} 
\includegraphics[scale=.9]{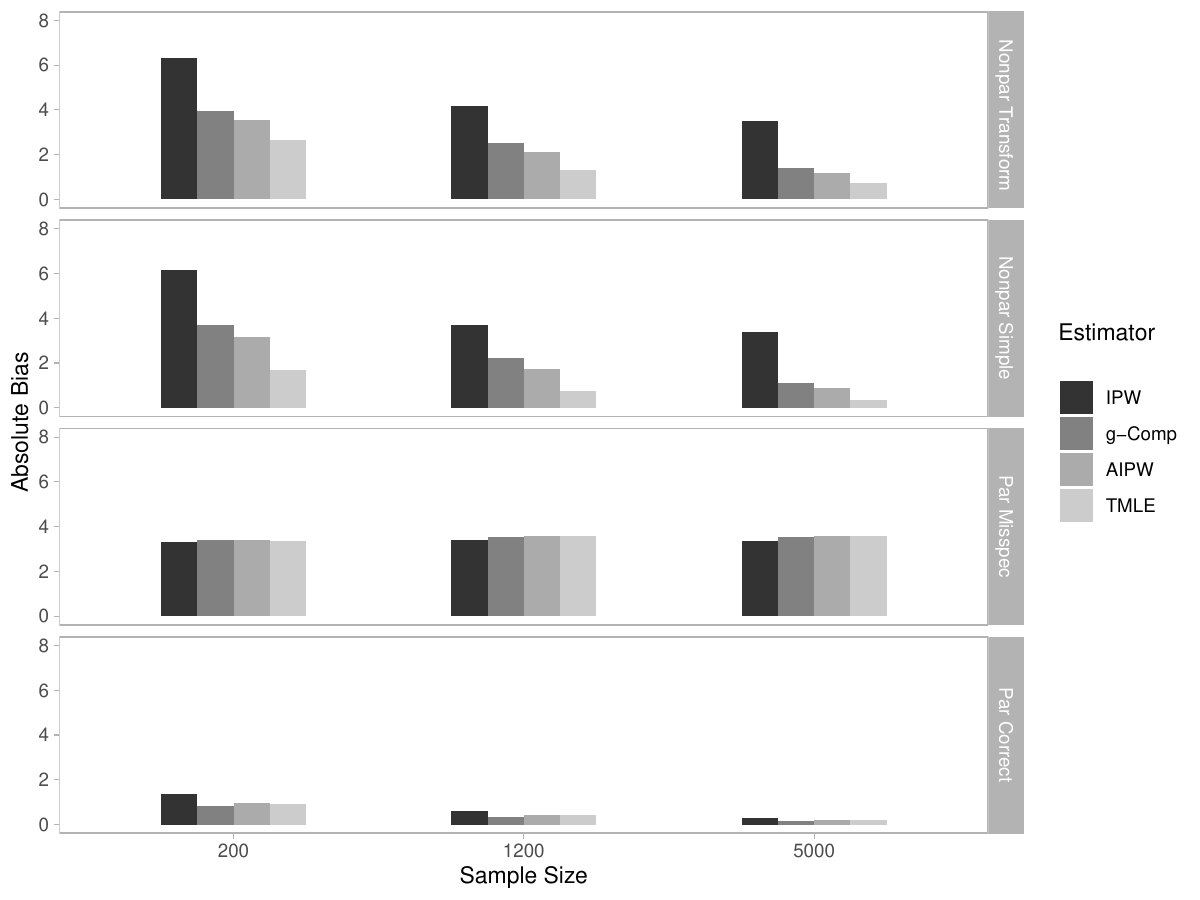}
\normalsize
\caption{Absolute bias of inverse probability weighted, g-computation, and doubly robust estimators for sample sizes of $N=200$, $N=1200$, and $N=5000$ when models for each estimator are specified parametrically (Par correct, Par misspec) using linear regression, and nonparametrically using a stacked generalization with random forests and extreme gradient boosting algorithms, and no sample splitting.} \label{F1}
\rule{\textwidth}{1pt}
\end{center}
\end{figure}

Figure \ref{F1} shows the estimated absolute bias across all sample sizes for all scenarios with the stacking algorithm that included random forests and extreme gradient boosting, and which did not use sample splitting. As expected, when using the correct parametric models, all methods are unbiased. In contrast, when the transformed confounders are used with parametric models (and thus parametric models are all mis-specified), all four estimators are subject to considerable bias which does not improve as the sample size increases (Figure \ref{F1}).

When models are fit nonparametrically using the simple confounders, IP-weighting displays considerable bias. G computation is also biased, but less than IP-weighting. In the nonparametric simple and complex settings (with transformed confounders), the bias decreases when doubly robust estimators are used  (Figure \ref{F1}). Generally, these results demonstrate what is expected from theory: the bias of singly robust estimators is larger than the bias of doubly robust estimators. Notably, in our simulation scenario under select sample sizes, the bias of the IP-weighted estimator under a nonparametric model with simple and transformed confounders is comparable to the bias of the misspecified parametric models (Figure \ref{F1}).

Table \ref{T1} shows the 95\% confidence interval coverage for each scenario. When correct parametric models were used, CI coverage was nominal, except for the robust variance estimator used for IP-weighting, which is known to be conservative.\cite{Hernan2001} When parametric models were fit with the transformed covariates (Parametric Misspecified), coverage dropped to 46\% or lower.

\begin{table}[ht]\rule{\textwidth}{1pt}
\caption{Confidence interval coverage$^\star$ for sample sizes of $N=200$, $N=1200$, and $N=5000$ obtained from parametric and nonparametric models under simple and complex confounding scenarios without sample splitting. Nonparametric estimation was based on a stacked generalization with random forests and extreme gradient boosting algorithms.\label{T1}}
\centering
\begin{tabular}{l|rrrr|rrrr}
 &  \multicolumn{3}{l}{{ Parametric True}} & & \multicolumn{3}{l}{{ Parametric Mispecified}} \\
$N$  & IPW & g-Comp & AIPW & TMLE & IPW & g-Comp & AIPW & TMLE \\ 
  \hline
      &&&&&&&&\\[-.5em]
   200 & 0.96 & 0.95 & 0.95 & 0.94 & 0.46 & 0.23 & 0.28 & 0.24 \\ 
  1200 & 0.98 & 0.93 & 0.94 & 0.94 & 0.01 & 0.00 & 0.00 & 0.00 \\ 
  5000 & 0.97 & 0.92 & 0.92 & 0.92 & 0.00 & 0.00 & 0.00 & 0.00 \\ 
	 \hline
      &&&&&&&&\\[-.5em]
      	 \hline
 & \multicolumn{3}{l}{{ Nonparametric Simple}} & & \multicolumn{3}{l}{{ Nonparametric Complex}} \\
$N$ &  IPW & g-Comp & AIPW & TMLE & IPW & g-Comp & AIPW & TMLE \\    
\hline
      &&&&&&&&\\[-.5em]
200  & 0.01 & NA & 0.02 & 0.22 & 0.00 & NA & 0.00 & 0.07 \\ 
1200 & 0.02 & NA & 0.00 & 0.24 & 0.01 & NA & 0.00 & 0.05 \\ 
5000 & 0.00 & NA & 0.02 & 0.29 & 0.00 & NA & 0.00 & 0.03 \\
      &&&&&&&&\\[-.5em]
   \hline
   \multicolumn{9}{p{12cm}}{{ $\star$ \footnotesize{Confidence interval coverage, defined as the proportion of 95\% confidence intervals that included the true value.}}}
\end{tabular}
\rule{\textwidth}{1pt}
\end{table}

The machine learning results presented in Table \ref{T1} represent version 1 of the stacked generalization when sample splitting was not used. When fit with machine learning algorithms, coverage for all estimators was well below the nominal threshold of 95\%. This was true for both singly and doubly robust approaches in both simple and transformed confounder settings (Table \ref{T1}).  

\begin{figure}[H]\rule{\textwidth}{1pt}
\begin{center} 
	\includegraphics[scale=.8]{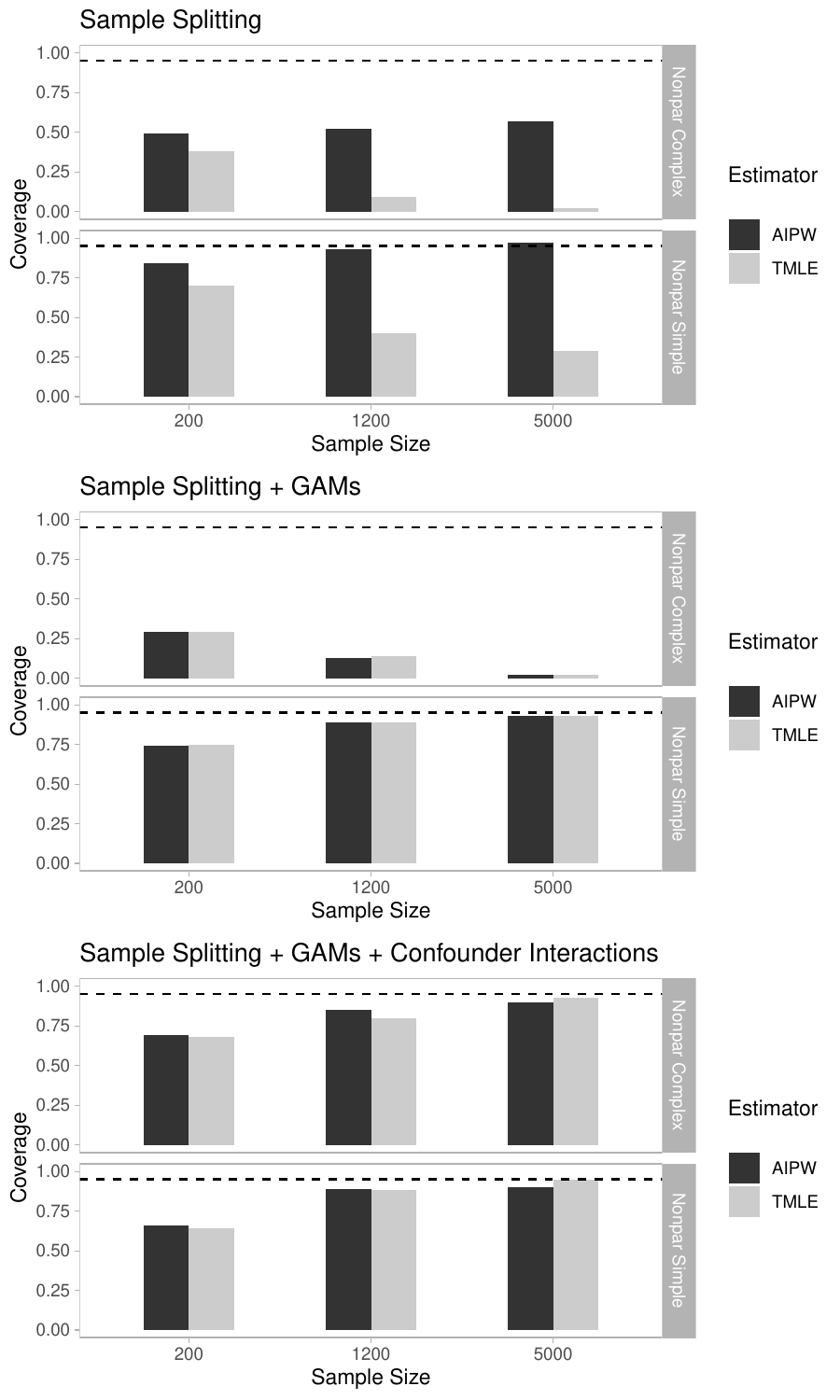}
\normalsize
\caption{Coverage of doubly robust estimators for sample sizes of $N=200$, $N=1200$, and $N=5000$ when models for each estimator are specified nonparametrically in the simple confounder and transformed confounder settings.} \label{F2}
\rule{\textwidth}{1pt}
\end{center}
\end{figure}

The poor performance of machine learning methods observed in Table \ref{T1} improved under the additional strategies explored. These results are presented in Figure \ref{F2}, which includes confidence interval coverage from scenarios in which: sample splitting, generalized additive models, and confounder interactions were used with the stacking algorithms and estimators. Indeed, the highest observed coverage was 29\% for TMLE in the simple confounder setting. In contrast, the lowest coverage in the simple confounder setting was 29\% for TMLE with sample splitting. When sample splitting was used, AIPW reached roughly nominal coverage rates in the simple confounder setting. Coverage improved in the transformed confounder setting with sample splitting, but did not reach nominal rates. 

When GAMs were combined with sample splitting, nominal coverage was attained in the simple confounder setting, but was still quite low for the transformed confounders. Coverage in the transformed confounder setting only attained nominal rates for AIPW and TMLE when sample splitting was combined with GAMs, and all confounder-confounder interactions were included in the models (Figure \ref{F2}).

\newpage

\section*{Discussion}

Both machine learning and doubly robust estimation are becoming increasingly popular, however the relation between them remains poorly understood. Here, we have shown how machine learning methods are biased when used with singly robust estimators such as inverse probability weighting or g computation (also known as marginal standardization). Performance, however, is greatly improved when used with doubly robust approaches, particularly with sample splitting and flexible regression methods.

Doubly robust estimators are often said to offer two chances to adjust for confounding or missing data. in fact, they offer additional protection against model misspecification. Model misspecification can occur for a number of reasons, including incorrect causal ordering of variables, incomplete confounder adjustment set, or incorrect functional form. This latter type of misspecification--incorrect functional form--is specifically what doubly robust estimators protect against. Indeed, excluding important confounders or including mediators that should not be adjusted for cannot be fixed with doubly robust methods alone.\cite{Keil2018}

A misspecified functional form can occur if the analyst fails to correctly account for the manner in which exposure and confounders relate to the outcome. For a generalized linear model, this would occur if chosen link function is not compatible with how the data were actually generated,\cite{Weisberg1994} if the analyst fails to account for curvilinear relations between the covariates and the outcome, or fails to include important exposure-confounder or confounder-confounder interactions. Unfortunately, in an observational study the true nature of these relations is typically not known.

Nonparametric techniques based on machine learning algorithms offer a degree of protection against each of these functional form assumptions. This feature has motivated a growing body of work in which data-adaptive methods are used to estimate parameters of interest. In particular, a number of authors have advocated for use of machine learning methods to estimate propensity scores,\cite{McCaffrey2013,Westreich2010c,Lee2010} or to mitigate against the strict parametric assumptions required by the g computation algorithm.\cite{Snowden2011,Westreich2015,Oulhote2019}

But, as we have shown, for singly robust estimators this protection may not always be worth the price. Under our chosen data generating mechanisms, implementing each estimator using correct parametric models resulted in unbiased estimation. However, when implemented nonparametrically using the correct set of confounders, both g computation and inverse probability weighting were biased, while both doubly robust approaches were less biased. These results align with other work on the use of machine learning methods with double robust estimators,\cite{Zivich2020,Chernozhukov2018,Kennedy2016}  and suggest that researchers should carefully weigh all considerations when using machine learning methods to estimate causal effects.

More specifically, our results suggest that when machine learning is used to quantify average treatment effects, researchers should employ the following techniques to maximize the performance of the estimation approach:

\begin{itemize}
	\item[1.] Use doubly robust estimation methods. These included augmented inverse probability weighted or targeted minimum loss-based estimation, or both. 
	\item[2.] Use sample splitting, also referred to as cross-fitting, double cross-fitting, or cross-validation, which improves estimation of standard errors and confidence interval coverage.
	\item[3.] Use a richly specified library of flexible regression, tree-based, gradient based, and other algorithms, that maximize the diversity of a given stacking algorithm. 
	\item[4.] Include first and higher order interactions between selected adjustment variables in a given stacking algorithm. Additionally, one may include other transformations (e.g., log, non-product interactions, or polynomial terms), as well as consider the use of screening algorithms that remove unnecessary variable transformations.
\end{itemize}

While our recommendations are general enough to be considered any time researchers seek to use machine learning methods when estimating causal effects, certain limitations of our simulation study should be taken into consideration. First, we did not focus our simulations on evaluating the relative performance of AIPW versus TMLE. Though are results might suggest that one or the other estimator performs better in certain settings, we would recommend against making such interpretations without a more in-depth exploration. Second, doubly robust-type methods have been developed for a wide variety of settings, including continuous\cite{Munoz2012,Kennedy2017b} and time-varying exposures,\cite{Kennedy2018} instrumental variables,\cite{Ogburn2015} mediation,\cite{Tchetgen2012c} and missing data.\cite{Long2012,Sun2018} Our simple simulation focused exclusively on using doubly robust methods to adjust for confounding. However, we do expect our findings would apply more generally.\cite{Kennedy2016} Finally, our simulations were very limited in that they explored a relatively simple data generating mechanism. Nevertheless, even under this simple data generating mechanism, we were only able to achieve low bias and nominal coverage only when sample splitting and flexible regression methods were used (for the simple confounder scenario), or when sample splitting, flexible regression, and confounder interactions were used (for the transformed confounder setting). We believe these findings should inform future analyses using machine learning methods with double robust estimators.

We have shown that, when used with singly robust approaches, nonparametric estimation techniques can yield suboptimal statistical properties. However, this can be ameliorated by using nonparametric methods with doubly robust estimators. In general, the choice between estimators should be motivated by their statistical properties, after one has selected an appropriate estimand that corresponds well to the research question at hand.\cite{Petersen2014} Taking full advantage of machine learning methods requires implementation of doubly robust estimation with sample splitting, the inclusion of flexible regression based methods into an appropriately selected meta-learner, and the inclusion of relevant covariate transformations/interactions.


\newpage

\footnotesize
\begin{thebibliography}{10}
\providecommand{\url}[1]{\texttt{#1}}
\providecommand{\urlprefix}{URL }
\providecommand{\bibinfo}[2]{#2}

\bibitem{Lee2010}
\bibinfo{author}{Lee BK}, \bibinfo{author}{Lessler J}, and
  \bibinfo{author}{Stuart EA}.
\newblock \bibinfo{title}{Improving propensity score weighting using machine
  learning}.
\newblock \emph{\bibinfo{journal}{Stat Med}}. \bibinfo{year}{2010};
  \textbf{\bibinfo{volume}{29}}:\bibinfo{pages}{337--346}.

\bibitem{Westreich2010c}
\bibinfo{author}{Westreich D}, \bibinfo{author}{Lessler J}, and
  \bibinfo{author}{Funk MJ}.
\newblock \bibinfo{title}{Propensity score estimation: neural networks, support
  vector machines, decision trees (cart), and meta-classifiers as alternatives
  to logistic regression}.
\newblock \emph{\bibinfo{journal}{J Clin Epidemiol}}. \bibinfo{year}{2010};
  \textbf{\bibinfo{volume}{63}}:\bibinfo{pages}{826 -- 833}.

\bibitem{Snowden2011}
\bibinfo{author}{Snowden JM}, \bibinfo{author}{Rose S}, and
  \bibinfo{author}{Mortimer KM}.
\newblock \bibinfo{title}{Implementation of g-computation on a simulated data
  set: Demonstration of a causal inference technique}.
\newblock \emph{\bibinfo{journal}{Am J Epidemiol}}. \bibinfo{year}{2011};
  \textbf{\bibinfo{volume}{173}}:\bibinfo{pages}{731--738}.

\bibitem{Oulhote2019}
\bibinfo{author}{Oulhote Y}, \bibinfo{author}{Coull B}, \bibinfo{author}{Bind
  MA} \emph{et~al.}
\newblock \bibinfo{title}{Joint and independent neurotoxic effects of early
  life exposures to a chemical mixture: A multi-pollutant approach combining
  ensemble learning and g-computation}.
\newblock \emph{\bibinfo{journal}{Environmental Epidemiology}}.
  \bibinfo{year}{2019}; \textbf{\bibinfo{volume}{3}}:\bibinfo{pages}{e063}.

\bibitem{Chernozhukov2018}
\bibinfo{author}{Chernozhukov V}, \bibinfo{author}{Chetverikov D},
  \bibinfo{author}{Demirer M} \emph{et~al.}
\newblock \bibinfo{title}{Double/debiased machine learning for treatment and
  structural parameters}.
\newblock \emph{\bibinfo{journal}{The Econometrics Journal}}.
  \bibinfo{year}{2018}; \textbf{\bibinfo{volume}{21}}:\bibinfo{pages}{C1--C68}.

\bibitem{Hastie2009}
\bibinfo{author}{Hastie T}, \bibinfo{author}{Tibshirani R}, and
  \bibinfo{author}{Friedman JH}.
\newblock \emph{\bibinfo{title}{The Elements of Statistical Learning: Data
  Mining, Inference, and Prediction}}.
\newblock \bibinfo{address}{New York, NY}: \bibinfo{publisher}{Springer}.
  \bibinfo{year}{2009}.

\bibitem{Naimi2020}
\bibinfo{author}{Naimi A}, \bibinfo{author}{Kennedy EH},
  \bibinfo{author}{Bodnar L}, \bibinfo{author}{EF S}, and \bibinfo{author}{SR
  C}.
\newblock \bibinfo{title}{Understanding and dealing with the curse of
  dimensionality}.
\newblock \emph{\bibinfo{journal}{Epidemiol}}. \bibinfo{year}{In Prep}; .

\bibitem{Robins1995b}
\bibinfo{author}{Robins J} and \bibinfo{author}{Rotnitzky A}.
\newblock \bibinfo{title}{Semiparametric efficiency in multivariate regression
  models with missing data}.
\newblock \emph{\bibinfo{journal}{JASA}}. \bibinfo{year}{1995};
  \textbf{\bibinfo{volume}{90}}:\bibinfo{pages}{122--9}.

\bibitem{Robins2001a}
\bibinfo{author}{Robins J} and \bibinfo{author}{Rotnitzky A}.
\newblock \bibinfo{title}{Comment: Inference for semiparametric models: Some
  questions and an answer}.
\newblock \emph{\bibinfo{journal}{Statistica Sinica}}. \bibinfo{year}{2001};
  \textbf{\bibinfo{volume}{11}}:\bibinfo{pages}{920--936}.

\bibitem{Bang2005}
\bibinfo{author}{Bang H} and \bibinfo{author}{Robins JM}.
\newblock \bibinfo{title}{Doubly robust estimation in missing data and causal
  inference models}.
\newblock \emph{\bibinfo{journal}{Biometrics}}. \bibinfo{year}{2005};
  \textbf{\bibinfo{volume}{61}}:\bibinfo{pages}{962---973}.

\bibitem{Rotnitzky2014}
\bibinfo{author}{Rotnitzky A} and \bibinfo{author}{Vansteelandt S}.
\newblock \bibinfo{title}{Double-robust methods}.
\newblock In: \bibinfo{editor}{Molenberghs G}, \bibinfo{editor}{Fitzmaurice G},
  \bibinfo{editor}{Kenward MG}, \bibinfo{editor}{Tsiatis A}, and
  \bibinfo{editor}{Verbeke G} (Eds.) \emph{\bibinfo{booktitle}{Handbook of
  Missing Data Methodology}}, chap.~\bibinfo{chapter}{9}.
  \bibinfo{publisher}{CRC Press}. \bibinfo{year}{2014};
  \bibinfo{pages}{185--209}.

\bibitem{Jonsson-Funk2011}
\bibinfo{author}{Jonsson-Funk M}, \bibinfo{author}{Westreich D},
  \bibinfo{author}{Wiesen C}, \bibinfo{author}{St\"{u}rmer T},
  \bibinfo{author}{Brookhart MA}, and \bibinfo{author}{Davidian M}.
\newblock \bibinfo{title}{Doubly robust estimation of causal effects}.
\newblock \emph{\bibinfo{journal}{Am J Epidemiol}}. \bibinfo{year}{2011};
  \textbf{\bibinfo{volume}{173}}:\bibinfo{pages}{761--767}.

\bibitem{vanderLaan2006}
\bibinfo{author}{van~der Laan MJ} and \bibinfo{author}{Rubin D}.
\newblock \bibinfo{title}{Targeted maximum likelihood learning}.
\newblock \emph{\bibinfo{journal}{Int J Biostat}}. \bibinfo{year}{2006};
  \textbf{\bibinfo{volume}{2}}:\bibinfo{pages}{Article 11}.

\bibitem{Kennedy2017}
\bibinfo{author}{Kennedy EH} and \bibinfo{author}{Balakrishnan S}.
\newblock \bibinfo{title}{{Discussion of ``Data-driven confounder selection via
  Markov and Bayesian networks'' by Jenny H\"{a}ggstr\"{o}m}}.
\newblock \emph{\bibinfo{journal}{Biometrics}}. \bibinfo{year}{2017};
  \textbf{\bibinfo{volume}{In Press}}.

\bibitem{Metropolis1949}
\bibinfo{author}{Metropolis N} and \bibinfo{author}{Ulam S}.
\newblock \bibinfo{title}{{The Monte Carlo method}}.
\newblock \emph{\bibinfo{journal}{J Am Stat Assoc}}. \bibinfo{year}{1949};
  \textbf{\bibinfo{volume}{44}}:\bibinfo{pages}{335--341}.

\bibitem{Greenland1999a}
\bibinfo{author}{Greenland S}, \bibinfo{author}{Pearl J}, and
  \bibinfo{author}{Robins J}.
\newblock \bibinfo{title}{Causal diagrams for epidemiological research}.
\newblock \emph{\bibinfo{journal}{Epidemiol}}. \bibinfo{year}{1999};
  \textbf{\bibinfo{volume}{10}}:\bibinfo{pages}{37--48}.

\bibitem{Robins2001}
\bibinfo{author}{Robins J}.
\newblock \bibinfo{title}{Data, design, and background knowledge in etiologic
  inference}.
\newblock \emph{\bibinfo{journal}{Epidemiol}}. \bibinfo{year}{2001};
  \textbf{\bibinfo{volume}{12}}:\bibinfo{pages}{313--320}.

\bibitem{Robins2009}
\bibinfo{author}{Robins JM} and \bibinfo{author}{Hern\'{a}n MA}.
\newblock \bibinfo{title}{Estimation of the causal effects of time-varying
  exposures}.
\newblock In: \bibinfo{editor}{Fitzmaurice G}, \bibinfo{editor}{Davidian M},
  \bibinfo{editor}{Verbeke G}, and \bibinfo{editor}{Molenberghs G} (Eds.)
  \emph{\bibinfo{booktitle}{Advances in Longitudinal Data Analysis}}.
  \bibinfo{address}{Boca Raton, FL}: \bibinfo{publisher}{Chapman \& Hall}.
  \bibinfo{year}{2009}; \bibinfo{pages}{553--599}.

\bibitem{Naimi2016b}
\bibinfo{author}{Naimi AI}, \bibinfo{author}{Cole SR}, and
  \bibinfo{author}{Kennedy EH}.
\newblock \bibinfo{title}{{An Introduction to G Methods}}.
\newblock \emph{\bibinfo{journal}{Int J Epidemiol}}. \bibinfo{year}{2016};
  \textbf{\bibinfo{volume}{In Press}}.

\bibitem{Hernan2006}
\bibinfo{author}{Hern\'{a}n MA} and \bibinfo{author}{Robins JM}.
\newblock \bibinfo{title}{Estimating causal effects from epidemiological data}.
\newblock \emph{\bibinfo{journal}{J Epidemiol Community Health}}.
  \bibinfo{year}{2006};
  \textbf{\bibinfo{volume}{60}}:\bibinfo{pages}{578--586}.

\bibitem{Daniel2018}
\bibinfo{author}{Daniel RM}.
\newblock \bibinfo{title}{Double robustness}.
\newblock In: \emph{\bibinfo{booktitle}{Wiley StatsRef: Statistics Reference
  Online}}. \bibinfo{publisher}{John Wiley \& Sons, Ltd}. \bibinfo{year}{2018};
  .

\bibitem{Rose2011}
\bibinfo{author}{Rose S} and \bibinfo{author}{van~der Laan MJ}.
\newblock \emph{\bibinfo{title}{Targeted learning: causal inference for
  observational and experimental data}}.
\newblock \bibinfo{address}{New York, NY}: \bibinfo{publisher}{Springer}.
  \bibinfo{year}{2011}.

\bibitem{Gruber2012}
\bibinfo{author}{Gruber S} and \bibinfo{author}{van~der Laan MJ}.
\newblock \bibinfo{title}{tmle: An r package for targeted maximum likelihood
  estimation}.
\newblock \emph{\bibinfo{journal}{Journal of Statistical Software}}.
  \bibinfo{year}{2012}; \textbf{\bibinfo{volume}{51}}:\bibinfo{pages}{1--35}.

\bibitem{Cole2013a}
\bibinfo{author}{Cole SR}, \bibinfo{author}{Chu H}, and
  \bibinfo{author}{Greenland S}.
\newblock \bibinfo{title}{Maximum likelihood, profile likelihood, and penalized
  likelihood: A primer.}
\newblock \emph{\bibinfo{journal}{Am J Epidemiol}}. \bibinfo{year}{2013};
  \textbf{\bibinfo{volume}{179}}:\bibinfo{pages}{252--260}.

\bibitem{Longford2008}
\bibinfo{author}{Longford N}.
\newblock \emph{\bibinfo{title}{Studying Human Populations: An Advanced Course
  in Statistics}}.
\newblock \bibinfo{address}{New York}: \bibinfo{publisher}{Springer}.
  \bibinfo{year}{2008}.

\bibitem{Rencher2000}
\bibinfo{author}{Rencher AC}.
\newblock \emph{\bibinfo{title}{{Linear Models in Statistics}}}.
\newblock \bibinfo{address}{New York}: \bibinfo{publisher}{Wiley}.
  \bibinfo{year}{2000}.

\bibitem{Box1976}
\bibinfo{author}{Box GEP}.
\newblock \bibinfo{title}{{Science and Statistics}}.
\newblock \emph{\bibinfo{journal}{JASA}}. \bibinfo{year}{1976};
  \textbf{\bibinfo{volume}{71}}:\bibinfo{pages}{791--99}.

\bibitem{Hernan2001}
\bibinfo{author}{Hern\'{a}n MA}, \bibinfo{author}{Brumback B}, and
  \bibinfo{author}{Robins JM}.
\newblock \bibinfo{title}{Marginal structural models to estimate the joint
  causal effect of nonrandomized treatments}.
\newblock \emph{\bibinfo{journal}{J Am Stat Assoc}}. \bibinfo{year}{2001};
  \textbf{\bibinfo{volume}{96}}:\bibinfo{pages}{440--448}.

\bibitem{Westreich2012a}
\bibinfo{author}{Westreich D}, \bibinfo{author}{Cole SR},
  \bibinfo{author}{Schisterman EF}, and \bibinfo{author}{Platt RW}.
\newblock \bibinfo{title}{{A simulation study of finite-sample properties of
  marginal structural Cox proportional hazards models}}.
\newblock \emph{\bibinfo{journal}{Stat Med}}. \bibinfo{year}{2012};
  \textbf{\bibinfo{volume}{31}}:\bibinfo{pages}{2098--2109}.

\bibitem{Kang2007}
\bibinfo{author}{Kang J} and \bibinfo{author}{JL S}.
\newblock \bibinfo{title}{Demystifying double robustness: A comparison of
  alternative strategies for estimating a population mean from incomplete
  data.}
\newblock \emph{\bibinfo{journal}{Stat Sci}}. \bibinfo{year}{2007};
  \textbf{\bibinfo{volume}{22}}:\bibinfo{pages}{523--539}.

\bibitem{vanderVaart2000}
\bibinfo{author}{van~der Vaart AW}.
\newblock \emph{\bibinfo{title}{Asymptotic statistics}}.
\newblock \bibinfo{address}{Cambridge}: \bibinfo{publisher}{Cambridge
  University Press}. \bibinfo{year}{2000}.

\bibitem{Gyorfi2002}
\bibinfo{author}{Gy\"{o}rfi L}, \bibinfo{author}{Kohler M},
  \bibinfo{author}{Krzyzak A}, and \bibinfo{author}{Walk H}.
\newblock \emph{\bibinfo{title}{A Distribution-Free Theory of Nonparametric
  Regression}}.
\newblock \bibinfo{address}{New York, NY}: \bibinfo{publisher}{Springer}.
  \bibinfo{year}{2002}.

\bibitem{Robins1997c}
\bibinfo{author}{Robins JM} and \bibinfo{author}{Ritov Y}.
\newblock \bibinfo{title}{Toward a curse of dimensionality appropriate (coda)
  asymptotic theory for semi-parametric models}.
\newblock \emph{\bibinfo{journal}{Stat Med}}. \bibinfo{year}{1997};
  \textbf{\bibinfo{volume}{16}}:\bibinfo{pages}{285--319}.

\bibitem{Wasserman2006}
\bibinfo{author}{Wasserman L}.
\newblock \emph{\bibinfo{title}{All of nonparametric statistics}}.
\newblock \bibinfo{address}{New York; London}: \bibinfo{publisher}{Springer}.
  \bibinfo{year}{2006}.

\bibitem{vanderLaan2007}
\bibinfo{author}{van~der Laan MJ}, \bibinfo{author}{Polley EC}, and
  \bibinfo{author}{Hubbard AE}.
\newblock \bibinfo{title}{Super learner}.
\newblock \emph{\bibinfo{journal}{Statistical Applications in Genetics and
  Molecular Biology}}. \bibinfo{year}{2007};
  \textbf{\bibinfo{volume}{6}}:\bibinfo{pages}{Article 25}.

\bibitem{Naimi2018}
\bibinfo{author}{Naimi AI} and \bibinfo{author}{Balzer LB}.
\newblock \bibinfo{title}{Stacked generalization: an introduction to super
  learning.}
\newblock \emph{\bibinfo{journal}{Eur J Epidemiol}}. \bibinfo{year}{2018};
  \textbf{\bibinfo{volume}{33}}:\bibinfo{pages}{459--464}.

\bibitem{Rinaldo2018}
\bibinfo{author}{Rinaldo A}, \bibinfo{author}{Wasserman L},
  \bibinfo{author}{G'Sell M}, and \bibinfo{author}{Lei J}.
\newblock \bibinfo{title}{Bootstrapping and sample splitting for
  high-dimensional, assumption-free inference}.
\newblock \emph{\bibinfo{journal}{https://arxivorg/abs/161105401}}.
  \bibinfo{year}{2018}; .

\bibitem{Zivich2020}
\bibinfo{author}{Zivich PN} and \bibinfo{author}{Breskin A}.
\newblock \bibinfo{title}{Machine learning for causal inference: on the use of
  cross-fit estimators}.
\newblock \emph{\bibinfo{journal}{arXiv:200410337}}. \bibinfo{year}{2020}; .

\bibitem{Keil2018}
\bibinfo{author}{Keil AP}, \bibinfo{author}{Mooney SJ},
  \bibinfo{author}{Jonsson~Funk M}, \bibinfo{author}{Cole SR},
  \bibinfo{author}{Edwards JK}, and \bibinfo{author}{Westreich D}.
\newblock \bibinfo{title}{Resolving an apparent paradox in doubly robust
  estimators.}
\newblock \emph{\bibinfo{journal}{Am J Epidemiol}}. \bibinfo{year}{2018};
  \textbf{\bibinfo{volume}{187}}:\bibinfo{pages}{891--892}.

\bibitem{Weisberg1994}
\bibinfo{author}{Weisberg S} and \bibinfo{author}{Welsh AH}.
\newblock \bibinfo{title}{Adapting for the missing link}.
\newblock \emph{\bibinfo{journal}{The Annals of Statistics}}.
  \bibinfo{year}{1994};
  \textbf{\bibinfo{volume}{22}}:\bibinfo{pages}{1674---1700}.

\bibitem{McCaffrey2013}
\bibinfo{author}{McCaffrey DF}, \bibinfo{author}{Griffin BA},
  \bibinfo{author}{Almirall D}, \bibinfo{author}{Slaughter ME},
  \bibinfo{author}{Ramchand R}, and \bibinfo{author}{Burgette LF}.
\newblock \bibinfo{title}{A tutorial on propensity score estimation for
  multiple treatments using generalized boosted models}.
\newblock \emph{\bibinfo{journal}{Stat Med}}. \bibinfo{year}{2013};
  \textbf{\bibinfo{volume}{32}}:\bibinfo{pages}{3388--3414}.

\bibitem{Westreich2015}
\bibinfo{author}{Westreich D}, \bibinfo{author}{Edwards JK},
  \bibinfo{author}{Cole SR}, \bibinfo{author}{Platt RW},
  \bibinfo{author}{Mumford SL}, and \bibinfo{author}{Schisterman EF}.
\newblock \bibinfo{title}{Imputation approaches for potential outcomes in
  causal inference}.
\newblock \emph{\bibinfo{journal}{International Journal of Epidemiology}}.
  \bibinfo{year}{2015}; :\bibinfo{pages}{Published ahead of print July 25,
  2015}.

\bibitem{Kennedy2016}
\bibinfo{author}{Kennedy EH}.
\newblock \bibinfo{title}{Semiparametric theory and empirical processes in
  causal inference}.
\newblock In: \bibinfo{editor}{He H}, \bibinfo{editor}{Wu P}, and
  \bibinfo{editor}{Chen DGD} (Eds.) \emph{\bibinfo{booktitle}{Statistical
  Causal Inferences and Their Applications in Public Health Research}}.
  \bibinfo{address}{Switzerland}: \bibinfo{publisher}{Springer International}.
  \bibinfo{year}{2016}; .

\bibitem{Munoz2012}
\bibinfo{author}{Munoz ID} and \bibinfo{author}{van~der Laan M}.
\newblock \bibinfo{title}{Population intervention causal effects based on
  stochastic interventions.}
\newblock \emph{\bibinfo{journal}{Biometrics}}. \bibinfo{year}{2012};
  \textbf{\bibinfo{volume}{68}}:\bibinfo{pages}{541--549}.

\bibitem{Kennedy2017b}
\bibinfo{author}{Kennedy EH}, \bibinfo{author}{Ma Z}, \bibinfo{author}{McHugh
  MD}, and \bibinfo{author}{Small DS}.
\newblock \bibinfo{title}{Non-parametric methods for doubly robust estimation
  of continuous treatment effects}.
\newblock \emph{\bibinfo{journal}{Journal of the Royal Statistical Society:
  Series B (Statistical Methodology)}}. \bibinfo{year}{2017};
  \textbf{\bibinfo{volume}{79}}:\bibinfo{pages}{1229--1245}.

\bibitem{Kennedy2018}
\bibinfo{author}{Kennedy EH}.
\newblock \bibinfo{title}{Nonparametric causal effects based on incremental
  propensity score interventions}.
\newblock \emph{\bibinfo{journal}{Journal of the American Statistical
  Association}}. \bibinfo{year}{2018}; :\bibinfo{pages}{1--12}.

\bibitem{Ogburn2015}
\bibinfo{author}{Ogburn EL}, \bibinfo{author}{Rotnitzky A}, and
  \bibinfo{author}{Robins JM}.
\newblock \bibinfo{title}{Doubly robust estimation of the local average
  treatment effect curve}.
\newblock \emph{\bibinfo{journal}{Journal of the Royal Statistical Society:
  Series B (Statistical Methodology)}}. \bibinfo{year}{2015};
  \textbf{\bibinfo{volume}{77}}:\bibinfo{pages}{373--396}.

\bibitem{Tchetgen2012c}
\bibinfo{author}{Tchetgen~Tchetgen EJ} and \bibinfo{author}{Shpitser I}.
\newblock \bibinfo{title}{Semiparametric theory for causal mediation analysis:
  Efficiency bounds, multiple robustness and sensitivity analysis}.
\newblock \emph{\bibinfo{journal}{Annals of Statistics}}. \bibinfo{year}{2012};
  \textbf{\bibinfo{volume}{40}}:\bibinfo{pages}{1816--1845}.

\bibitem{Long2012}
\bibinfo{author}{Long Q}, \bibinfo{author}{Hsu CH}, and \bibinfo{author}{Li Y}.
\newblock \bibinfo{title}{Doubly robust nonparametric multiple imputation for
  ignorable missing data.}
\newblock \emph{\bibinfo{journal}{Stat Sin}}. \bibinfo{year}{2012};
  \textbf{\bibinfo{volume}{22}}:\bibinfo{pages}{149--172}.

\bibitem{Sun2018}
\bibinfo{author}{Sun B} and \bibinfo{author}{Tchetgen~Tchetgen EJ}.
\newblock \bibinfo{title}{On inverse probability weighting for nonmonotone
  missing at random data.}
\newblock \emph{\bibinfo{journal}{J Am Stat Assoc}}. \bibinfo{year}{2018};
  \textbf{\bibinfo{volume}{113}}:\bibinfo{pages}{369--379}.

\bibitem{Petersen2014}
\bibinfo{author}{Petersen ML} and \bibinfo{author}{van~der Laan MJ}.
\newblock \bibinfo{title}{Causal models and learning from data: integrating
  causal modeling and statistical estimation}.
\newblock \emph{\bibinfo{journal}{Epidemiol}}. \bibinfo{year}{2014};
  \textbf{\bibinfo{volume}{25}}:\bibinfo{pages}{418--426}.

\end{thebibliography}
\end{document}